\documentclass[aip,jcp,reprint,amsmath,amssymb,floatfix,citeautoscript,nofootinbib]{revtex4-2}
\usepackage{times}
\usepackage{mathptmx}
\DeclareMathAlphabet{\mathcal}{OMS}{cmsy}{m}{n} 
\usepackage[T1]{fontenc}
\usepackage{algorithm}
\usepackage{algpseudocode}
\usepackage{bm}
\usepackage{amsmath}
\usepackage{amssymb}
\usepackage{xcolor}
\usepackage{graphicx}
\usepackage{hyperref}
\hypersetup{
    colorlinks=true,
    linkcolor=blue,
    urlcolor=blue,
    citecolor=blue,
}
\usepackage{cleveref}
\usepackage{braket}
\usepackage[version=4]{mhchem}

\newcommand{\me}{\mathrm{e}}

\definecolor{myred}{rgb}{0.8,0,0}

\begin{document}

    \title{Unphysical Solutions in Coupled-Cluster-Based Random Phase Approximation and How to Avoid Them}

    \author{Ruiheng Song}
    \affiliation{Department of Chemistry and Biochemistry, University of Maryland, College Park, MD 20742}

    \author{Xiliang Gong}
    \affiliation{Department of Chemistry and Biochemistry, University of Maryland, College Park, MD 20742}

    \author{Hong-Zhou Ye}
    \email{hzye@umd.edu}
    \affiliation{Department of Chemistry and Biochemistry, University of Maryland, College Park, MD 20742}
    \affiliation{Institute for Physical Science and Technology, University of Maryland, College Park, MD 20742}
    \date{\today}

    \begin{abstract}

        The direct ring coupled-cluster doubles (drCCD)-based random phase approximation (RPA) has provided an attractive framework for the development and application of RPA-related methods.
        However, a potential unphysical solution issue recently reported by Rekkedal and co-workers (\textit{J.~Chem.~Phys}.~\textbf{139}, 081101, 2013) has raised significant concerns regarding the general applicability of coupled-cluster-based RPA, particularly in small-gap systems where RPA is anticipated to outperform commonly employed second-order perturbation theory.
        In this work, we elucidate the underlying origin of the multi-solution issue in drCCD and develop both a practical criterion for validating drCCD solutions and improved preconditioners based on level shifting and regularized MP2 methods for stabilizing the iterative solution of the drCCD equation.
        We demonstrate the robustness and effectiveness of our approach through representative systems---including molecules with stretched bonds, large conjugated systems, and metallic clusters---where standard drCCD iteration encounters convergence difficulties.
        Furthermore, we extend our approach to various recently developed reduced-scaling drCCD-based RPA methods, thereby establishing a foundation for their stable application to large-scale problems.
        The extension of our approach to RPA with exchange, quasiparticle RPA, and particle-particle RPA is also discussed.

    \end{abstract}

    \maketitle


    The random phase approximation~\cite{Bohm53PR,GellMann57PR} (RPA) is a widely employed post-mean-field method in computational materials science~\cite{Ren12JMS,Chen17ARPC}, for its ability to improve upon semilocal and hybrid Kohn-Sham density functional theory~\cite{Hohenberg64PR,Kohn65PR} in systems involving non-covalent interactions~\cite{Lu09PRL,DelBen13JCTC,Klimes16JCP,Zen18PNAS,Yao21JPCL,Pham23JCP} and those of metallic character~\cite{Schimka10NM,Ren09PRB,Olsen11PRL,Rohlfin08PRL,Ma11PRB,Kim12PRB,Schmidt18JPCC,Sheldon24JCTC}.
    The RPA correlation energy can be evaluated through several distinct formulations, including the plasmon formula~\cite{Bohm53PR}, the generator coordinate method~\cite{Jancovici64NP}, the adiabatic-connection formulation~\cite{Langreth77PRB}, and the direct ring coupled-cluster doubles (drCCD) approach~\cite{Scuseria08JCP}.
    This work focuses on the drCCD-based RPA framework, which has facilitated the development of various low-scaling RPA algorithms~\cite{Kallay15JCP,Liang25JCTC} and beyond-RPA methods~\cite{Gruneis09JCP,Mussard14JCTC,Hehn15JCP}.
    However, a potential multi-solution issue recently identified by Rekkedal and co-workers~\cite{Rekkedal13JCP} has raised substantial concerns regarding the general reliability of drCCD-based RPA.
    Specifically, these authors discovered an unphysical drCCD solution with energy significantly lower than the expected RPA energy for \ce{H2} in a stretched geometry when employing the standard iterative algorithm to solve the drCCD equation.
    The extent to which this multi-solution behavior affects the practical applicability of drCCD has remained an unresolved question.

    In this work, we investigate the fundamental origin of unphysical solutions in drCCD and develop algorithms for robustly converging to the desired physical solution.
    Specifically, we demonstrate that
    (i) the total number of solutions to a drCCD equation equals the dimension of the full configuration space within a given orbital basis;
    (ii) all solutions except one are unphysical, with energies lower than the physical solution by a partial sum of the RPA excitation energies;
    (iii) a necessary and sufficient condition for validating the physicality of a drCCD solution with amplitude $T$ is that $\lambda_{\textrm{max}}(T^{\dagger}T)$---the maximum eigenvalue of $T^{\dagger} T$---be less than unity;
    (iv) the commonly employed iterative procedure for solving the drCCD equation, based on an MP2-style preconditioner, either converges to an unphysical solution or fails to converge entirely for small-gap systems, as exemplified by molecules with stretched bonds, large conjugated systems, and metallic clusters;
    (v) improving the preconditioner through either level shifting or $\sigma$-regularization enables robust convergence to the physical solution; and
    (vi) these improved preconditioners can be adapted to stabilize various reduced-scaling drCCD-based RPA methods, thereby establishing a foundation for reliable RPA calculations on a large scale.

    In what follows, we assume a mean-field reference with integer occupation and canonical orbitals.
    We will use $i,j,k$ for occupied orbitals, $a,b,c$ for virtual orbitals, $p,q,r$ for unspecified orbitals, and $\varepsilon$ for the corresponding orbital energies.
    The number of occupied and virtual orbitals are $N_{\textrm{occ}}$ and $N_{\textrm{vir}}$, respectively.
    The number of single particle-hole excitations is $N_{\textrm{ov}} = N_{\textrm{occ}} \times N_{\textrm{vir}}$.
    For simplicity, the following discussion focuses on direct particle-hole RPA (hereafter referred to as RPA) with a stable mean-field reference.
    Extensions to RPA with exchange (RPAx, also known as full RPA), quasi-particle RPA~\cite{Scuseria13JCP}, particle-particle RPA~\cite{Scuseria13JCP,Peng13JCP}, and RPA with an unstable mean-field reference are provided in the Supporting Information.
    All numerical calculations presented in this work were performed using a developer version of the PySCF package~\cite{Sun18WIRCMS,Sun20JCP}, with additional computational details provided in the Supporting Information.

    We begin with a brief review of the connection between RPA and drCCD.
    The plasmon formula for the RPA correlation energy is given by
    \begin{equation}    \label{eq:Ec_RPA}
        E_{\textrm{c}}^{\textrm{RPA}}
            = \frac{1}{2} \textrm{Tr} \{\Omega - A\}
            = \frac{1}{2} \sum_{n=1}^{N_{\textrm{ov}}} \omega_n - \omega_{n}^{\textrm{TDA}}
    \end{equation}
    where $\Omega = \textrm{diag}\{\omega_1, \omega_2, \cdots, \omega_{N_{\textrm{ov}}}\}$ contains the positive eigenvalues of the RPA eigenvalue equation
    \begin{equation}    \label{eq:RPA_eigeqn}
        \begin{bmatrix}
            A & B \\
            -B^* & -A^*
        \end{bmatrix}
        \begin{bmatrix}
            X & Y^* \\
            Y & X^*
        \end{bmatrix}
            = \begin{bmatrix}
                X & Y^* \\
                Y & X^*
            \end{bmatrix}
            \begin{bmatrix}
                \Omega & 0 \\
                0 & -\Omega
            \end{bmatrix}
    \end{equation}
    with
    \begin{equation}    \label{eq:AB_definition}
    \begin{split}
        A_{iajb}
            &= (\varepsilon_a - \varepsilon_i) \delta_{ij}\delta_{ab} + (ia|bj),  \\
        B_{iajb}
            &= (ia|jb),
    \end{split}
    \end{equation}
    and $\Omega^{\textrm{TDA}} = \textrm{diag}\{\omega_1^{\textrm{TDA}}, \omega_2^{\textrm{TDA}}, \cdots, \omega_{N_{\textrm{ov}}}^{\textrm{TDA}}\}$ represents the Tamm-Dancoff approximation~\cite{Hirata99CPL} (TDA) to RPA, defined by
    \begin{equation}
        A X^{\textrm{TDA}}
            = X^{\textrm{TDA}} \Omega^{\textrm{TDA}}.
    \end{equation}

    Scuseria and co-workers demonstrated that $E_{\textrm{c}}^{\textrm{RPA}}$ can alternatively be computed through drCCD~\cite{Scuseria08JCP},
    \begin{equation}    \label{eq:Ec_drCCD}
        E_{\textrm{c}}^{\textrm{RPA}}
            = E_{\textrm{c}}^{\textrm{drCCD}}(T)
            = \frac{1}{2} \textrm{Tr}~BT
    \end{equation}
    where
    \begin{equation}    \label{eq:T_physical}
        T = Y X^{-1}
    \end{equation}
    is a solution to the drCCD residual equation,
    \begin{equation}    \label{eq:drCCD_eqn}
        R(T)
            = B^* + A^* T + T A + T B T
            = 0.
    \end{equation}
    This alternative formulation of RPA has facilitated the development of reduced-scaling RPA algorithms~\cite{Kallay15JCP,Liang25JCTC} and beyond-RPA methods~\cite{Gruneis09JCP,Mussard14JCTC,Hehn15JCP}.
    When using anti-symmetrized electron repulsion integrals in \cref{eq:AB_definition}, namely $\braket{ib||aj}$ for $A$ and $\braket{ij||ab}$ for $B$, \cref{eq:Ec_RPA} yields the RPAx energy, which is reproduced by the corresponding ring CCD theory~\cite{Scuseria08JCP}, also given by \cref{eq:drCCD_eqn} but with anti-symmetrized integrals.

    Extending the proof by Scuseria and co-workers~\cite{Scuseria08JCP}, one can demonstrate that the drCCD equation (\ref{eq:drCCD_eqn}) admits a complete family of solutions, given by
    \begin{equation}    \label{eq:T_unphysical}
        \tilde{T}_{\bm{\eta}}
            = \tilde{Y}_{\bm{\eta}} \tilde{X}_{\bm{\eta}}^{-1}
    \end{equation}
    where $\eta_n \in \{+1, -1\}$ for $n = 1, 2, \cdots, N_{\textrm{ov}}$ specifies whether the $n$-th column of $\tilde{X}_{\bm{\eta}}$ and $\tilde{Y}_{\bm{\eta}}$ is taken from the RPA eigenvectors corresponding to the positive eigenvalues (i.e.,~$X$ and $Y$) or the negative eigenvalues (i.e.,~$Y^*$ and $X^*$).
    The corresponding drCCD correlation energy is
    \begin{equation}    \label{eq:Ec_drCCD_unphysical}
        E_{\textrm{c}}^{\textrm{drCCD}}[\tilde{T}_{\bm{\eta}}]
            = \frac{1}{2} \sum_{n=1}^{N_{\textrm{ov}}} \eta_n \omega_n - \omega_{n}^{\textrm{TDA}}
            = E_{\textrm{c}}^{\textrm{RPA}} - \sum_{n=1}^{N_{\textrm{ov}}} \delta_{\eta_{n},-1} \omega_{n},
    \end{equation}
    which is always lower than $E_{\textrm{c}}^{\textrm{RPA}}$ unless $\eta_n = +1$ for all $n$, in which case $\tilde{T}_{\bm{\eta}}$ reduces to $T$ in \cref{eq:T_physical} and the correct $E_{\textrm{c}}^{\textrm{RPA}}$ is recovered.
    For this reason, \cref{eq:T_physical} is the \textit{physical} solution of drCCD, while all other solutions $\tilde{T}_{\bm{\eta}}$ containing one or more $-1$ values in $\bm{\eta}$ are termed \textit{unphysical} due to their lower energies.
    The total number of solutions given by \cref{eq:T_unphysical}, encompassing both physical and unphysical solutions, is $2^{N_{\textrm{ov}}}$, which coincides with the dimension of the full configuration space.

    A natural question arises: does \cref{eq:T_unphysical} enumerate \textit{all} solutions to the drCCD equation?
    When the RPA eigenvalue equation possesses real and non-degenerate eigenvalues, the drCCD equation corresponds to the continuous-time algebraic Riccati equation (CARE) of a stable system in control theory, which can admit at most $2^{N_{\textrm{ov}}}$ solutions~\cite{Willems71IEEETAC,Kucera73Kybernetika}.
    We therefore conclude that when the mean-field reference is stable and the RPA spectrum is non-degenerate, the corresponding drCCD equation has exactly $2^{N_{\textrm{ov}}}$ solutions.
    When degeneracy occurs in the RPA spectrum, the eigenvectors within a degenerate subspace cannot be uniquely determined, as any unitary rotations mixing them leave \cref{eq:RPA_eigeqn} invariant.
    However, the physical solution (\ref{eq:T_physical}) remains uniquely defined because $\mathbf{X}$ and $\mathbf{Y}$ undergo the same unitary rotation, thereby canceling each other's effect.
    An unphysical solution is also uniquely defined provided that $\eta_n$ values corresponding to a degenerate subspace are identical (this encompasses the physical solution as a special case). When some degenerate subspace contains both positive and negative $\eta_n$ values, different unitary rotations can be independently applied to the degenerate positive and negative eigenvectors, resulting in infinitely many unphysical solutions.

    \begin{table}[!h]
        \centering
        \caption{$E_{\textrm{c}}^{\textrm{drCCD}}$ and its deviation from $E_{\textrm{c}}^{\textrm{dRPA}}$ for the physical solution and six representative unphysical solutions of a water molecule at its equilibrium geometry in the cc-pVDZ basis set, generated from the RPA eigenvectors using \cref{eq:T_unphysical} with the corresponding $\bm{\eta}$ vectors specified in the first column.
        The energies of the unphysical solutions are systematically lower than $E_{\textrm{c}}^{\textrm{dRPA}}$ by the corresponding sum of RPA eigenvalues.
        The final column displays the maximum eigenvalue of $T^{\dagger} T$, which is less than unity for the physical solution and exceeds unity for all unphysical solutions.
        The reference state is the PBE ground state.
        All energies are given in Hartree units.}
        \label{tab:water}
        \begin{tabular}{llll}
            \hline\hline
            State
                & $E_{\textrm{c}}^{\textrm{drCCD}}$
                & $E_{\textrm{c}}^{\textrm{drCCD}} - E_{\textrm{c}}^{\textrm{dRPA}}$
                & $\lambda_{\textrm{max}}(T^{\dagger}T)$ \\
            \hline
            $\bm{\eta} = +\bm{1}$ & $-0.308280$ & $\phantom{-}0.000000$ & $0.068$ \\
            $\eta_1 = -1$ & $-0.589456$ & $-0.281176 = -\omega_1$ & $632.5$ \\
            $\eta_2 = -1$ & $-0.657723$ & $-0.349443 = -\omega_2$ & $2218.3$ \\
            $\eta_3 = -1$ & $-0.675253$ & $-0.366973 = -\omega_3$ & $822.5$ \\
            $\eta_{1,2} = -1$ & $-0.938898$ & $-0.630618 = -(\omega_1+\omega_2)$ & $2218.3$ \\
            $\eta_{1,3} = -1$ & $-0.956428$ & $-0.648148 = -(\omega_1+\omega_3)$ & $822.5$ \\
            $\eta_{1,2,3} = -1$ & $-1.305871$ & $-0.997591 = -(\omega_1+\omega_2+\omega_3)$ & $2218.3$ \\
            \hline
        \end{tabular}
    \end{table}

    The multiple-solution issue of drCCD is illustrated numerically for a water molecule in \cref{tab:water}, where $E_{\textrm{c}}^{\textrm{drCCD}}$ and its deviation from $E_{\textrm{c}}^{\textrm{dRPA}}$ are presented for the physical solution and six representative unphysical solutions, all generated using \cref{eq:T_unphysical} with different choices of $\bm{\eta}$.
    As expected, the physical solution reproduces the correct RPA correlation energy, while all unphysical solutions exhibit energies lower than $E_{\textrm{c}}^{\textrm{RPA}}$ by the corresponding sum of RPA eigenvalues.

    In practice, the drCCD equation (\ref{eq:drCCD_eqn}) is solved iteratively starting from an initial guess to circumvent the high computational cost of fully diagonalizing the RPA matrix.
    The presence of unphysical solutions can either impede convergence to the physical solution or lead to convergence toward an unphysical one.
    We first address the fundamental question of whether one can determine if a given drCCD solution is physical without resorting to solving the RPA eigenvalue equation.
    To this end, we introduce a necessary and sufficient criterion for identifying the physical drCCD solution:
    \begin{equation}    \label{eq:lambda_max_square}
        \lambda_{\textrm{max}}(T^{\dagger}T)
            < 1,
    \end{equation}
    where $\lambda_{\textrm{max}}(T^{\dagger}T)$ denotes the maximum eigenvalue of $T^{\dagger} T$.
    This condition is numerically validated for the water molecule in \cref{tab:water}, where all unphysical solutions violate \cref{eq:lambda_max_square}.
    Since $T^{\dagger} T$ is positive-semidefinite, $\lambda_{\textrm{max}}(T^{\dagger}T)$ can be efficiently computed with $O(N_{\textrm{ov}}^2) \sim O(N^4)$ computational scaling using the Davidson algorithm~\cite{Davidson75JCP} [which can be further reduced to $O(N_{\textrm{ov}}N_{\textrm{aux}}) \sim O(N^3)$ using the factorized drCCD equation (\ref{eq:drCCD_eqn_U}) discussed below].

    The proof of \cref{eq:lambda_max_square} begins by recognizing that the RPA eigenvectors define a bosonic Bogoliubov transformation and therefore possess the following Bloch-Messiah decomposition~\cite{Ring80Book}:
    \begin{equation}    \label{eq:Bloch_Messiah}
        \begin{bmatrix}
            X & Y^* \\
            Y & X^*
        \end{bmatrix}
            = \begin{bmatrix}
                U & 0 \\
                0 & U^*
            \end{bmatrix}
            \begin{bmatrix}
                \bar{X} & \bar{Y} \\
                \bar{Y} & \bar{X}
            \end{bmatrix}
            \begin{bmatrix}
                V & 0 \\
                0 & V^*
            \end{bmatrix}^{\dagger}
    \end{equation}
    where $U$ and $V$ are unitary matrices and
    \begin{equation}    \label{eq:barX_barY}
    \begin{split}
        \bar{X}
            &= \textrm{diag}\{\cosh(\theta_1), \cosh(\theta_2), \cdots, \cosh(\theta_{N_{\textrm{ov}}})\},  \\
        \bar{Y}
            &= \textrm{diag}\{\sinh(\theta_1), \sinh(\theta_2), \cdots, \sinh(\theta_{N_{\textrm{ov}}})\}
    \end{split}
    \end{equation}
    for some $\theta_{n} \in \mathbb{R}$ with $n = 1, 2, \cdots, N_{\textrm{ov}}$.
    The hyperbolic sine and cosine functions in \cref{eq:barX_barY} are a natural result of the normalization condition of the RPA eigenvectors, $X^{\dagger} X - Y^{\dagger} Y = 1$.
    \Cref{eq:Bloch_Messiah} enables us to rewrite the family of drCCD solutions (\ref{eq:T_unphysical}) in the Autonne-Takagi decomposition form~\cite{Houde24CJP}:
    \begin{equation}    \label{eq:T_unphysical_AT}
        T_{\bm{\eta}}
            = U^* \Lambda_{\bm{\eta}} U^{\dagger}
    \end{equation}
    where $\Lambda_{\bm{\eta}}$ is a diagonal matrix with elements
    \begin{equation}
        \lambda_n
            = \left\{
            \begin{split}
                \tanh(\theta_n),&\quad{}\eta_n = +1 \\
                \coth(\theta_n),&\quad{}\eta_n = -1
            \end{split}
            \right.
    \end{equation}
    Using \cref{eq:T_unphysical_AT}, we obtain the desired eigenvalue decomposition of $T^{\dagger} T$:
    \begin{equation}    \label{eq:TdaggerT_unphysical_AT}
        T_{\bm{\eta}}^{\dagger} T_{\bm{\eta}}
            = U \Lambda_{\bm{\eta}}^2 U^{\dagger}.
    \end{equation}
    For a physical solution with $\eta_n = +1$ for all $n$, we have $\lambda_{\textrm{max}}(T^{\dagger} T) = \max_n \tanh^2(\theta_n) \in [0,1)$, while for an unphysical solution containing one or more $\eta_n = -1$, we obtain $\lambda_{\textrm{max}}(T^{\dagger} T) = \max_{n: \eta_n = -1} \coth^2(\theta_n) \in (1,\infty)$.
    This establishes \cref{eq:lambda_max_square} as a necessary and sufficient condition for identifying the physical solution.
    For calculations employing real orbitals, $T$ is a real-symmetric matrix whose eigenvalue decomposition is given by \cref{eq:T_unphysical_AT}.
    In this case, we can alternatively verify whether $|\lambda(T)|_{\textrm{max}}$---the maximum absolute eigenvalue of $T$---is less than unity.

    While condition (\ref{eq:lambda_max_square}) enables validation of the physicality of a drCCD solution \textit{a posteriori}, in practice it is more desirable to prevent convergence to an unphysical solution \textit{a priori}.
    This motivation leads us to examine the convergence behavior of commonly employed iterative algorithms for solving the drCCD equation.
    To facilitate the following discussion, we rewrite \cref{eq:drCCD_eqn} as
    \begin{equation}    \label{eq:drCCD_eqn_Delta_form}
        R(T)
            = \Delta \circ T + B^* + L^* T + T L + T B T = 0
    \end{equation}
    where $L_{ia,jb} = (ia|bj)$, $\Delta_{ia,jb} = \varepsilon_a + \varepsilon_b - \varepsilon_i - \varepsilon_j$, and ``$\circ$'' denotes the element-wise product.
    Starting with an initial guess $T_0$, the amplitudes are iteratively updated until convergence by taking an approximate Newton step:
    \begin{equation}    \label{eq:T_update}
        T_{n+1}
            = T_{n} - P(T_n) \circ R(T_n),
    \end{equation}
    where $P_{ia,jb}$ represents a diagonal preconditioner.
    Without loss of generality, we set $T_{-1} = 0$ to generate an initial guess $T_0 = P(0) \circ B^*$ that is consistent with the employed preconditioner.
    The iterative process is typically accelerated using the direct inversion of the iterative subspace (DIIS) algorithm~\cite{Pulay80CPL}.

    \begin{figure*}[!t]
        \centering
        \includegraphics[width=6.6in]{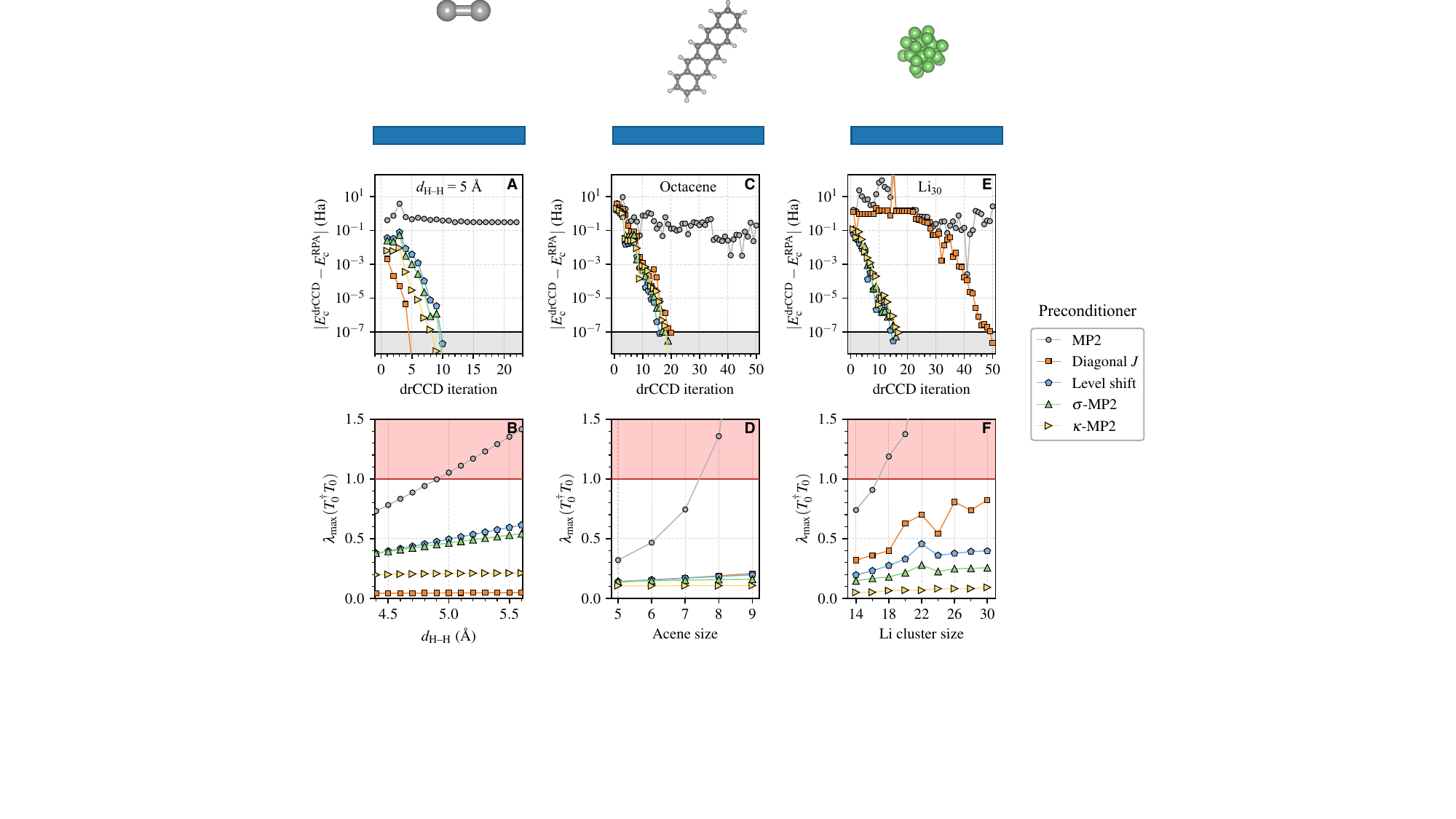}
        \caption{(A,C,E) Convergence of $E_{\textrm{c}}^{\textrm{drCCD}}$ using different preconditioners for solving the drCCD equation (\ref{eq:drCCD_eqn_Delta_form}) for (A) \ce{H2} with a bond length of $5~\textrm{\AA}$, (C) octacene, and (E) \ce{Li_{30}}.
        The gray shaded region indicates energy errors below $10^{-7}$~Ha.
        (B,D,F) $\lambda_{\textrm{max}}(T_0^{\dagger}T_0)$ of initial amplitudes $T_0$ generated by different preconditioners for (B) \ce{H2} with increasing bond length, (D) linear acenes of increasing size (quantified by the number of benzene rings), and (F) lithium clusters of increasing size.
        The red shaded region indicates $\lambda_{\textrm{max}}(T_0^{\dagger}T_0) > 1$.
        The level-shift, $\sigma$-MP2, and $\kappa$-MP2 preconditioners employ $\eta = 0.1$~Ha, $\sigma = 0.2$~Ha, and $\kappa = 0.2$~Ha, respectively.
        The \ce{H2} calculations use a Hartree-Fock reference, while the acene and lithium cluster calculations use a PBE reference.
        The cc-pVDZ/cc-pVDZ-JKFIT basis sets are employed for all calculations in conjunction with the frozen-core approximation.}
        \label{fig:precond_conv}
    \end{figure*}

    The most widely used MP2-style preconditioner,
    \begin{equation}    \label{eq:P_mp2}
        P_{ia,jb}^{\textrm{MP2}}
            = \Delta_{iajb}^{-1}
    \end{equation}
    is known to cause drCCD convergence difficulties for systems with small energy gaps.
    To the best of our knowledge, this issue was first reported by Rekkedal and co-workers when studying \ce{H2} dissociation~\cite{Rekkedal13JCP}.
    In \cref{fig:precond_conv}A, we reproduce their observations by demonstrating that the MP2 preconditioner leads to an unphysical drCCD solution for \ce{H2} with a bond length of $5~\textrm{\AA}$.
    This unphysical solution corresponds to $\eta_1 = -1$ in \cref{eq:T_unphysical}, with $E_{\textrm{c}}^{\textrm{drCCD}} = -0.445187$~Ha being lower than $E_{\textrm{c}}^{\textrm{RPA}} = -0.135110$~Ha by precisely the lowest RPA eigenvalue, $\omega_1 = 0.310077$~Ha.
    We have also verified that $\lambda_{\textrm{max}}(T^{\dagger}T) = 4.45 > 1$ for this solution.

    To gain insight into this unexpected convergence behavior, we present in \cref{fig:precond_conv}B the quantity $\lambda_{\textrm{max}}(T_0^{\dagger}T_0)$ for the initial MP2 amplitudes $T_0$ as a function of the \ce{H2} bond length.
    We observe that as the bond is stretched, $\lambda_{\textrm{max}}(T_0^{\dagger}T_0)$ increases monotonically and eventually enters the unphysical regime ($\lambda_{\textrm{max}} > 1$) for $d_{\textrm{H--H}} \gtrapprox 4.9~\textrm{\AA}$.
    This observation suggests that $\lambda_{\textrm{max}}(T_0^{\dagger}T_0)$ serves as a potential indicator for convergence toward an unphysical solution.

    Beyond bond dissociation, we have found that the MP2 preconditioner generally results in convergence difficulties for drCCD in small-gap systems.
    This behavior is demonstrated for linear acenes and lithium clusters in \cref{fig:precond_conv}, which serve as representative examples of large conjugated systems and metals, respectively.
    In \cref{fig:precond_conv}C and E, we show that the MP2 preconditioner produces large energy oscillations for octacene and \ce{Li_{30}}, failing to converge to a drCCD solution within $50$ cycles in both cases.
    Examination of $\lambda_{\textrm{max}}^2(T_0^{\dagger} T_0)$ for a series of linear acenes and lithium clusters of increasing size, shown in \cref{fig:precond_conv}D and F, reveals that both octacene and \ce{Li_{30}} exhibit $\lambda_{\textrm{max}}^2(T_0^{\dagger} T_0) \gg 1$, which is consistent with the \ce{H2} example discussed above.

    We now turn to examining several alternative preconditioners that can potentially stabilize the drCCD iteration compared to the MP2-based approach.
    Rekkedal and co-workers~\cite{Rekkedal13JCP} proposed a preconditioner equivalent to the inverse of the Jacobian matrix diagonal, which we term the diagonal-$J$ preconditioner:
    \begin{equation}    \label{eq:P_diagJ}
        P_{ia,jb}^{\textrm{Diagonal-}J}(T)
            = \left[
                \Delta_{ia,jb} + (L+TK)_{ia,ia} + (L+KT)_{jb,jb}
            \right]^{-1}.
    \end{equation}
    This preconditioner was demonstrated to successfully generate the physical solution in \ce{H2} dissociation, a result we have reproduced through our own calculations of \ce{H2} shown in \cref{fig:precond_conv}A.
    In \cref{fig:precond_conv}C and E, we demonstrate that the diagonal-$J$ preconditioner also yields the physical solution for octacene and \ce{Li_{30}}, although in the latter case, convergence is significantly slower and requires nearly $50$ iterations.
    The initial values of $\lambda_{\textrm{max}}(T_0^{\dagger}T_0)$ presented in \cref{fig:precond_conv}B, D, and F for these three systems using the diagonal-$J$ preconditioner provide insight into the different convergence rates observed:
    for both \ce{H2} and linear acenes, the initial $\lambda_{\textrm{max}}(T_0^{\dagger} T_0) \ll 1$, indicating that the initial guess lies close to the physical solution, while for lithium clusters, the initial $\lambda_{\textrm{max}}(T_0^{\dagger} T_0)$ exhibits noticeable growth with cluster size and approaches unity for \ce{Li_{30}}, which accounts for the slower convergence in this system.

    The MP2 preconditioner (\ref{eq:P_mp2}) can also be modified in several ways.
    A conceptually simple approach is applying a uniform shift $\eta > 0$ to the energy denominator, yielding the level-shift~\cite{Saunders73IJQC} preconditioner:
    \begin{equation}    \label{eq:P_level_shift}
        P_{ia,jb}^{\textrm{Level-shift}}(\eta)
            = (\Delta_{ia,jb} + \eta)^{-1}.
    \end{equation}
    A family of preconditioners can also be derived from various regularized MP2 methods~\cite{Shee21JPCL}.
    In this work, we consider two such preconditioners: one inspired by $\sigma$-MP2~\cite{Shee21JPCL,Evangelista14JCP},
    \begin{equation}    \label{eq:P_sigma_mp2}
        P^{\sigma\textrm{-MP2}}_{ia,jb}(\sigma)
            = \frac{1 - \me^{-\Delta_{ia,jb}/\sigma}}{\Delta_{ia,jb}},
    \end{equation}
    and another based on $\kappa$-MP2~\cite{Lee18JCTC},
    \begin{equation}    \label{eq:P_kappa_mp2}
        P^{\kappa\textrm{-MP2}}_{ia,jb}(\kappa)
            = \frac{\left(1 - \me^{-\Delta_{ia,jb}/\kappa}\right)^2}{\Delta_{ia,jb}}.
    \end{equation}
    Note that our definition of $\sigma$ and $\kappa$ is the reciprocal of their literature definitions.
    This choice ensures that all three regularization parameters ($\eta$, $\sigma$, and $\kappa$) possess units of energy, thereby facilitating their comparison.

    Ideally, the values of these parameters must be chosen sufficiently large to stabilize the calculation against unphysical solutions at early stages, yet sufficiently small to ensure rapid convergence at later stages.
    For the $\kappa$-MP2 preconditioner, determining an appropriate value of $\kappa$ to balance stability and efficiency can be challenging for small-gap systems, because $P_{ia,jb}^{\kappa\textrm{-MP2}}$ vanishes linearly with $\Delta_{ia,jb}$ as the latter approaches zero, significantly slowing down amplitude updates for particle-hole channels near the gap.
    The level-shift and $\sigma$-MP2 preconditioners are affected by the same issue, albeit to a lesser extent due to their different asymptotic behaviors in the limit of vanishing $\Delta$:
    \(P_{ia,jb}^{\textrm{Level-shift}} \to \eta^{-1}\) and \(P_{ia,jb}^{\sigma\textrm{-MP2}} \to \sigma^{-1}\), both approaching a positive constant.

    To address this challenge, we propose a two-stage algorithm where modified MP2 preconditioners are applied only during the early pre-convergence stage, followed by employing the bare MP2 preconditioner for final convergence.
    The transition between stages can be triggered by monitoring the error at each iteration.
    In this work, we employ an energy-based criterion:
    \begin{equation}    \label{eq:two_stage_transition}
        |E_{n} - E_{n-1}| < \eta_{\textrm{preconv}}.
    \end{equation}
    Our numerical experiments indicate that a relatively large value of $\eta_{\textrm{preconv}} = 0.1$~Ha effectively balances stability and convergence rate for all cases tested.
    Using this two-stage algorithm, we found reasonably optimal parameter values for the three modified MP2 preconditioners: $\eta = 0.1$~Ha and $\sigma = \kappa = 0.2$~Ha, which are employed throughout this work.

    In \cref{fig:precond_conv}, we present results for the three systems studied above using the three modified MP2 preconditioners (\ref{eq:P_level_shift}--\ref{eq:P_kappa_mp2}) with the two-stage algorithm.
    From \cref{fig:precond_conv}A, C, and E, we observe that all three preconditioners successfully stabilize the drCCD iteration for all three systems, matching the performance of the diagonal-$J$ preconditioner for stretched \ce{H2} and octacene while demonstrating significant improvement over the latter for \ce{Li_{30}}.
    This observation is consistent with the smaller values of $\lambda_{\textrm{max}}(T_0^{\dagger}T_0)$ exhibited by these three preconditioners for the lithium clusters (\cref{fig:precond_conv}F), which explain the much faster error decay in the first few iterations compared to using the diagonal-$J$ preconditioner.
    To emphasize the importance of the two-stage algorithm, we demonstrate in fig.~S1 that without employing it, all three preconditioners require more iterations to converge for \ce{Li_{30}} (which has a small gap of $0.0063$~Ha or $0.17$~eV), with the $\kappa$-MP2 preconditioner being most severely affected and failing to achieve convergence after $50$ cycles.
    Given their ease of implementation and robustness, we recommend all three modified MP2 preconditioners (\ref{eq:P_level_shift}--\ref{eq:P_kappa_mp2}) in conjunction with the two-stage algorithm for routine use in stabilizing the iterative solution of the drCCD equation.

    Finally, we discuss the implementation of the improved preconditioners introduced above for various reduced-scaling drCCD-based RPA methods.
    The domain-based localized pair natural orbital (DLPNO)-based RPA~\cite{Liang25JCTC}, recently introduced by one of the authors, solves the drCCD equation using the DLPNO approximation~\cite{Neese09JCPa,Neese09JCPb}.
    This method preserves the structure of the drCCD equation and therefore permits straightforward application of all preconditioners discussed above to stabilize the iteration.
    By contrast, the local natural orbital (LNO)-based RPA~\cite{Kallay15JCP} introduced by K\'{a}llay relies on a factorized drCCD equation, which requires additional consideration.
    This method begins by factorizing the drCCD equation (\ref{eq:drCCD_eqn_Delta_form}) as follows:
    \begin{equation}    \label{eq:drCCD_eqn_factorized}
        R(T)
            = \Delta \circ T + (J^* + T J) (J^* + T J)^{\top}
            = 0
    \end{equation}
    where $J$ is the density fitting factor for the electron-repulsion integrals:
    \begin{equation}
    \begin{split}
        K_{ia,jb}
            &= \sum_{x}^{N_{\textrm{aux}}} J_{iax} J_{jbx},  \\
        L_{ia,jb}
            &= \sum_{x}^{N_{\textrm{aux}}} J_{iax} J^*_{jbx},
    \end{split}
    \end{equation}
    where $x$ indexes a set of $N_{\textrm{aux}}$ auxiliary basis functions.
    Defining $U = J^* + TJ$ and employing \cref{eq:drCCD_eqn_factorized}, the drCCD iteration (\ref{eq:T_update}) can be rewritten as
    \begin{equation}    \label{eq:drCCD_eqn_U_intermediate}
        U_{n+1}
            = U_{n} - [P_n \circ \Delta \circ T_n] J - [ P_n \circ (U_n U_n^{\top})] J
    \end{equation}
    where $P_n$ denotes the preconditioner for the $n$-th iteration.
    Setting $P_n = P^{\textrm{MP2}}$ enables complete elimination of $T$ from \cref{eq:drCCD_eqn_U_intermediate}, yielding a factorized drCCD equation for $U$:
    \begin{equation}    \label{eq:drCCD_eqn_U}
        U_{n+1}
            = J^{*} - [P^{\textrm{MP2}} \circ (U_n U_n^{\top})] J.
    \end{equation}
    As written, \cref{eq:drCCD_eqn_U} exhibits $O(N^5)$ cost scaling.
    K\'{a}llay proposed further factorizing the positive-definite MP2 preconditioner using Cholesky decomposition:
    \begin{equation}    \label{eq:P_mp2_CD}
        P^{\textrm{MP2}}_{ia,jb}
            = \sum_{w}^{N_{\textrm{CD}}} \tau^{\textrm{MP2}}_{iaw} \tau^{\textrm{MP2}}_{jbw},
    \end{equation}
    which allows rewriting \cref{eq:drCCD_eqn_U} as
    \begin{equation}    \label{eq:drCCD_eqn_U_CD}
        U_{n+1}
            = J^{*}
            - \sum_{w}^{N_{\textrm{CD}}}
            (\tau_{w}^{\textrm{MP2}} \circ U_n)
            (\tau_{w}^{\textrm{MP2}} \circ U_n)^{\top} J.
    \end{equation}
    \Cref{eq:drCCD_eqn_U_CD} can be implemented with $O(N^4)$ computational scaling assuming $N_{\textrm{CD}}$ does not scale with system size.
    Combining \cref{eq:drCCD_eqn_U_CD} with the LNO approximation~\cite{Rolik11JCP,Rolik13JCP} results in a linear-scaling LNO-RPA algorithm~\cite{Kallay15JCP}.

    The preceding derivation reveals that the factorized drCCD iteration (\ref{eq:drCCD_eqn_U}) is equivalent to the full drCCD iteration (\ref{eq:T_update}) with a MP2 preconditioner and therefore suffers from similar convergence issues shown in \cref{fig:precond_conv}.
    While improved preconditioners cannot be directly applied to \cref{eq:drCCD_eqn_U_intermediate} because they prevent eliminating $T$ from the equation, our two-stage algorithm remains applicable.
    Specifically, a stabilizing preconditioner replaces $P^{\textrm{MP2}}$ in \cref{eq:drCCD_eqn_U} during pre-convergence, followed by $P^{\textrm{MP2}}$ for final convergence.
    The same energy-based criterion (\ref{eq:two_stage_transition}) can be used to determine the transition point.
    Furthermore, provided that the stabilizing preconditioner is positive-definite and amenable to Cholesky factorization [as in \cref{eq:P_mp2_CD}], our two-stage algorithm can be implemented with $O(N^4)$ scaling via \cref{eq:drCCD_eqn_U_CD} and accelerated using LNO.
    The three recommended preconditioners (\ref{eq:P_level_shift}--\ref{eq:P_kappa_mp2}) satisfy this condition, but the diagonal-$J$ preconditioner (\ref{eq:P_diagJ}) does not, as its $T$-dependent terms (which can be rewritten in terms of $U$) may violate positive-definiteness.

    \begin{figure}[!tbh]
        \centering
        \includegraphics[width=3.3in]{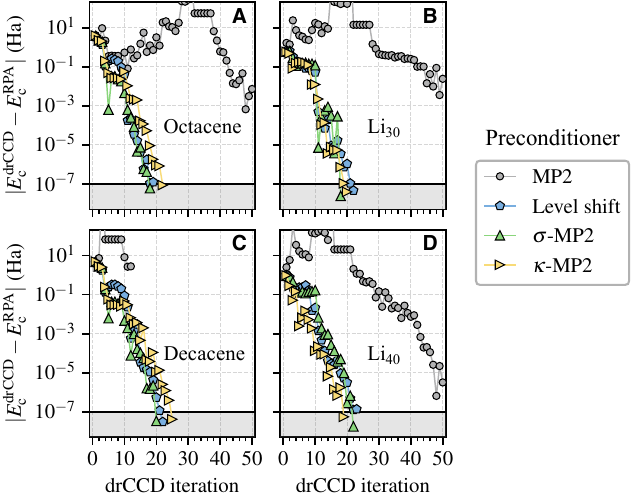}
        \caption{Convergence of $E_{\textrm{c}}^{\textrm{drCCD}}$ using different preconditioners for solving the factorized drCCD equation (\ref{eq:drCCD_eqn_U}) for (A) octacene, (B) \ce{Li_{30}}, (C) decacene, and (D) \ce{Li_{40}}.
        The gray shaded region indicates energy errors below $10^{-7}$~Ha.
        The level-shift, $\sigma$-MP2, and $\kappa$-MP2 preconditioners employ $\eta = 0.1$~Ha, $\sigma = 0.2$~Ha, and $\kappa = 0.2$~Ha, respectively.
        A PBE reference is used for all calculations in conjunction with the cc-pVDZ/cc-pVDZ-JKFIT basis sets and the frozen-core approximation.}
        \label{fig:quartic_conv}
    \end{figure}

    We implemented the quartic-scaling factorized drCCD equation (\ref{eq:drCCD_eqn_U_CD}) with the two-stage algorithm.
    In \cref{fig:quartic_conv}A and B, we demonstrate that all three modified MP2 preconditioners leads to rapid convergence to the physical solution of factorized drCCD for octacene and \ce{Li_{30}}, thus reproducing the results shown in \cref{fig:precond_conv}C and E for the full drCCD equation.
    The $O(N^4)$ computational scaling of the factorized drCCD method enables us to examine the convergence behavior for larger systems, as demonstrated in \cref{fig:quartic_conv}C and D for decacene and \ce{Li_{40}}, respectively.
    In both cases, all three preconditioners converge rapidly to the desired physical solution, consistent with their performance in smaller systems and representing a significant improvement over the standard MP2 preconditioner, which fails catastrophically for decacene due to severe linear dependencies in DIIS and converges very slowly for \ce{Li_{40}}.


    In conclusion, we have addressed the unphysical solution issue of drCCD-based RPA by establishing a practical criterion for validating drCCD solutions and developing a family of preconditioners that stabilize the drCCD iteration to ensure robust convergence to the physical solution.
    Our approach is general and also applicable to other variants of RPA as discussed in the Supporting Information.
    The effectiveness of our approach has been demonstrated using representative systems that pose convergence challenges for the standard drCCD iteration, including molecules with stretched bonds, large conjugated systems, and metallic clusters.
    Through the two-stage algorithm, we have shown that preconditioners based on level shifting and regularized MP2 methods enable rapid convergence to the desired physical drCCD solution in all systems studied.
    We have also demonstrated that our approach is fully compatible with various recently developed reduced-scaling RPA methods based on drCCD, thereby establishing a foundation for robust RPA calculations on a large scale.

    \section*{Supporting Information}

    See Supporting Information for (i) computational details, (ii) extension to other flavors of RPA and RPA with an unstable mean-field reference, (iii) optimized molecular structures, (iv) raw data presented in \cref{fig:precond_conv,fig:quartic_conv}, and (v) drCCD convergence without using the two-stage algorithm.

    \section*{Conflict of interest}
    The authors declare no competing conflicts of interest.

    \section*{Acknowledgments}

    This work was supported by the startup funds from the University of Maryland, College Park.
    We acknowledge computing resources provided by the Division of Information Technology at the University of Maryland, College Park.

    \bibliography{refs}

\end{document}


\title{Supporting Information for Unphysical Solutions in Coupled-Cluster-Based Random Phase Approximation and How to Avoid Them}

    \author{Ruiheng Song}
    \affiliation{Department of Chemistry and Biochemistry, University of Maryland, College Park, MD 20742}

    \author{Xiliang Gong}
    \affiliation{Department of Chemistry and Biochemistry, University of Maryland, College Park, MD 20742}

    \author{Hong-Zhou Ye}
    \email{hzye@umd.edu}
    \affiliation{Department of Chemistry and Biochemistry, University of Maryland, College Park, MD 20742}
    \affiliation{Institute for Physical Science and Technology, University of Maryland, College Park, MD 20742}


    \maketitle

    \vspace{3em}

    Note: figures, tables, and equations in the main text will be referred to as Fig.~\fakeref{M1}, Table~\fakeref{M2}, and Eq.~\fakeref{M3}.

    \section{Computational details}

    All calculations reported in this work using RPA and drCCD are performed using a developer version of the PySCF code~\cite{Sun18WIRCMS,Sun20JCP} with Libcint~\cite{Sun15JCC} for integral evaluation.
    The mean-field calculations are performed at zero-temperature with integer occupation.
    The reference RPA correlation energy is calculated using AC-RPA as implemented in PySCF with $60$ quadrature points for the imaginary frequency integration.
    The cc-pVDZ basis sets~\cite{Dunning89JCP} and the associated cc-pVDZ-JKFIT auxiliary basis sets~\cite{Weigend02PCCP} are used for both the mean-field and correlated calculations, along with the frozen-core approximation for the latter.
    Molecular structures optimized at B3LYP~\cite{Becke93JCP}+D3~\cite{Grimme10JCP}/def2-SVP~\cite{Weigend05PCCP} level using ORCA~\cite{Neese12WIRCMS,Neese18WIRCMS,Neese20JCP} can be found below in \cref{sec:supp_data}.
    For the factorized drCCD, the Cholesky decomposition (Eq.~\fakeref{M28}) vectors reproduce the matrix elements of the preconditioner to a precision better than $10^{-9}~\textrm{Ha}^{-1}$.
    For the iterative solution of both full and factorized drCCD equations, the maximum DIIS space is set to $6$.
    The iteration is deemed converged if both the energy change between two iterations drops below $10^{-7}$~Ha and the amplitude change ($T$ for full drCCD and $U$ for factorized drCCD) drops below $10^{-6}$ (a.u.).

    \section{Extension to other flavors of RPA}
    \label{sec:extension_other_flavor}

    The discussion in the main text for direct particle-hole RPA (ph-RPA) relies only on the mathematical properties of the RPA eigenvalue equation (Eq.~\fakeref{M2}), which is a symplectic eigenvalue equation.
    As a consequence, all results developed in the main text apply immediately to other flavors of RPA that can be formulated using a symplectic eigenvalue equation.
    This includes ph-RPA with exchange (commonly referred to as RPAx or full RPA), particle-particle RPA~\cite{Scuseria13JCP,Peng13JCP} (pp-RPA), and quasi-particle RPA~\cite{Scuseria13JCP} (qp-RPA), to name a few.
    Our discussion below for qp and pp-RPA follows closely Ref~\onlinecite{Scuseria13JCP}.

    \subsection{RPAx}

    The RPAx equation is the same as Eq.~\fakeref{M2} except that $A$ and $B$ contain anti-symmetrized electron repulsion integrals.
    The RPAx correlation energy can be reproduced by the physical solution of ring CCD, whose amplitude $T$ also satisfies the condition $\lambda_{\textrm{max}}(T^{\dagger} T) < 1$.


    \subsection{qp-RPA}

    The qp-RPA is RPA for a Hartree-Fock-Bogoliubov state~\cite{Scuseria13JCP}.
    The qp-RPA eigenvalue equation is also formally the same as Eq.~\fakeref{M2},
    \begin{equation}    \label{eq:qpRPA_eqn}
    \begin{bmatrix}
        \mathcal{A} & \mathcal{B} \\
        -\mathcal{B}^* & -\mathcal{A}^*
    \end{bmatrix}
    \begin{bmatrix}
        \mathcal{X} & \mathcal{Y}^* \\
        \mathcal{Y} & \mathcal{X}^*
    \end{bmatrix}
    =
    \begin{bmatrix}
        \mathcal{X} & \mathcal{Y}^* \\
        \mathcal{Y} & \mathcal{X}^*
    \end{bmatrix}
    \begin{bmatrix}
        \Omega & 0 \\
        0 & -\Omega^*
    \end{bmatrix},
    \end{equation}
    with $\mathcal{A}$ and $\mathcal{B}$ matrices given by
    \begin{equation}    \label{eq:AB_qp_RPA}
    \begin{split}
        A_{pq,rs} &= \braket{[\alpha_q \alpha_p,[H,\alpha_r^\dagger \alpha_s^\dagger]]}, \\
        B_{pq,rs} &= \braket{[\alpha_q \alpha_p,[H,\alpha_s \alpha_r]]},
    \end{split}
    \end{equation}
    where $\alpha_{p}$ and $\alpha_{p}^{\dagger}$ are quasi-particle annihilation and creation operators, respectively.
    All matrices, $\mathcal{A}$, $\mathcal{B}$, $\mathcal{X}$, $\mathcal{Y}$, $\Omega$, are of size $N_{\textrm{orb}} \times N_{\textrm{orb}}$ with $N_{\textrm{orb}} = N_{\textrm{occ}} + N_{\textrm{vir}}$.
    Under the assumption that the underlying HFB state is stable, the qp-RPA eigenvalues are all real, and a physical qp-RPA correlation energy is given by the plasmonic formula,
    \begin{equation}
        E_{\textrm{c}}^{\textrm{qp-RPA}}
            = \frac{1}{2} \textrm{Tr}\,\left(
                \Omega - \mathcal{A}
            \right)
            = \frac{1}{2} \textrm{Tr}\, \mathcal{B} \mathcal{T}
    \end{equation}
    where in the second equality, we introduced the physical solution
    \begin{equation}
        \mathcal{T}
            = \mathcal{Y} \mathcal{X}^{-1}
    \end{equation}
    to the Ricatti equation corresponding to the qp-RPA problem,
    \begin{equation}
        \mathcal{B}^* + \mathcal{A}^* \mathcal{T} + \mathcal{T} \mathcal{A}
        + \mathcal{T} \mathcal{B} \mathcal{T}
            = 0.
    \end{equation}
    The physical solution $\mathcal{T}$ satisfies the same condition, $\lambda_{\textrm{max}}(\mathcal{T}^{\dagger} \mathcal{T}) < 1$.

    \subsection{pp-RPA}

    When qp-RPA is applied to a stable HF (rather than HFB) state, the $\mathcal{A}$ and $\mathcal{B}$ matrices have the following block structure,
    \begin{equation}
        \mathcal{A}
            = \begin{bmatrix}
                \mathcal{A}_{\textrm{oo,oo}} & 0 & 0 \\
                0 & \mathcal{A}_{\textrm{ov,ov}} & 0 \\
                0 & 0 & \mathcal{A}_{\textrm{vv,vv}}
            \end{bmatrix},
        \quad{}
        \mathcal{B}
            = \begin{bmatrix}
                0 & 0 & \mathcal{B}_{\textrm{oo,vv}} \\
                0 & \mathcal{B}_{\textrm{ov,ov}} & 0 \\
                \mathcal{B}_{\textrm{vv,oo}} & 0 & 0
            \end{bmatrix}
    \end{equation}
    where $\mathcal{A}_{\textrm{ov,ov}}$ and $\mathcal{B}_{\textrm{ov,ov}}$ are the $A$ and $B$ matrices in ph-RPA (Eq.~\fakeref{M3}).
    Using the notation from Ref~\onlinecite{Scuseria13JCP},
    \begin{equation}
    \begin{split}
        \mathcal{A}_{\textrm{oo,oo}}
            = D,
        \qquad{}
        &D_{ij,kl}
            = -(\varepsilon_{i}+\varepsilon_{j}) \delta_{ik} \delta_{jl} + \braket{kl||ij}, \\
        \mathcal{A}_{\textrm{vv,vv}}
            = C,
        \qquad{}
        &C_{ab,cd}
            = (\varepsilon_{a}+\varepsilon_{b}) \delta_{ac} \delta_{bd} + \braket{ab||cd},  \\
        \mathcal{B}_{\textrm{oo,vv}}
            = \bar{B},
        \qquad{}
        &\bar{B}_{ij,ab}
            = \braket{ab||ij}.
    \end{split}
    \end{equation}
    Note that both $C$ and $D$ are Hermitian matrices.
    This block structure leads to three independent eigenvalue problem: the ov-ov blocks give the ph-RPA presented in the main text, while the other blocks give pp-RPA,
    \begin{equation}    \label{eq:ppRPA_eqn}
        \begin{bmatrix}
        C & -\bar{B} \\
        \bar{B}^\dagger & -D^*
        \end{bmatrix}
        \begin{bmatrix}
        X_1 & Y_2 \\
        Y_1 & X_2
        \end{bmatrix}
        =
        \begin{bmatrix}
        X_1 & Y_2 \\
        Y_1 & X_2
        \end{bmatrix}
        \begin{bmatrix}
        \Omega_1 & 0 \\
        0 & \Omega_2
        \end{bmatrix}
    \end{equation}
    and its particle-hole conjugate, the hole-hole RPA (hh-RPA),
    \begin{equation}    \label{eq:hhRPA_eqn}
        \begin{bmatrix}
            D & -\bar{B}^{\top} \\
            \bar{B}^* & -C^*
        \end{bmatrix}
        \begin{bmatrix}
            X_2^* & Y_1^* \\
            Y_2^* & X_1^*
        \end{bmatrix}
            = \begin{bmatrix}
                X_2^* & Y_1^* \\
                Y_2^* & X_1^*
            \end{bmatrix}
            \begin{bmatrix}
                -\Omega_2 & 0 \\
                0 & -\Omega_1
            \end{bmatrix},
    \end{equation}
    where $\Omega_1$ is positive and $\Omega_2$ is negative for a stable HF reference.
    The pp-RPA and hh-RPA have the same correlation energy
    \begin{equation}
        E_{\textrm{c}}^{\textrm{pp-RPA}}
            = \textrm{Tr}\,(\Omega_1 - C)
            = \textrm{Tr}\,(-\Omega_2 - D)
            = E_{\textrm{c}}^{\textrm{hh-RPA}}
    \end{equation}
    which is reproduced by the physical solution
    \begin{equation}
        T_1
            = -Y_1 X_1^{-1}
            = -(Y_2 X_2^{-1})^{\dagger}
            = T_2^{\dagger}
    \end{equation}
    to the ladder CCD equation
    \begin{equation}
        \bar{B}^\dagger + D^* T_1 + T_1 C + T_1 \bar{B} T_1 = 0.
    \end{equation}
    To derive the condition for the physical ladder CCD solution, we note that while neither pp-RPA (\ref{eq:ppRPA_eqn}) nor hh-RPA (\ref{eq:hhRPA_eqn}) corresponds to a symplectic eigenvalue equation, the combined problem
    \begin{equation}    \label{eq:pphhRPA_eqn}
        \begin{bmatrix}
            D & 0 & 0 & -\bar{B}^T \\
            0 & C & -\bar{B} & 0 \\
            0 & \bar{B}^\dagger & -D^* &0 \\
            \bar{B}^* & 0 & 0 & -C^*
        \end{bmatrix}
        \begin{bmatrix}
            X_2^* & 0 & 0 & Y_1^* \\
            0 & X_1 & Y_2 & 0 \\
            0 & Y_1 & X_2 &0 \\
            Y_2^* & 0 & 0 & X_1^*
        \end{bmatrix}
            = \begin{bmatrix}
                X_2^* & 0 & 0 & Y_1^* \\
                0 & X_1 & Y_2 & 0 \\
                0 & Y_1 & X_2 &0 \\
                Y_2^* & 0 & 0 & X_1^*
            \end{bmatrix}
            \begin{bmatrix}
                -\Omega_2 & 0 & 0 & 0 \\
                0 & \Omega_1 & 0 & 0 \\
                0 & 0 & \Omega_2 &0 \\
                0 & 0 & 0 & -\Omega_1
            \end{bmatrix}
    \end{equation}
    gives a symplectic eigenvalue equation.
    The plasmonic correlation energy of this pp+hh-RPA theory is equal to both the pp-RPA and the hh-RPA correlation energy,
    \begin{equation}
        E_{\textrm{cc}}^{\textrm{pp+hh-RPA}}
            = \frac{1}{2} \textrm{Tr}\,(\Omega_1 - \Omega_2 - C - D)
            = \underbrace{\textrm{Tr}\,(\Omega_1 - C)}_{E_{\textrm{cc}}^{\textrm{pp-RPA}}}
            = \underbrace{\textrm{Tr}\,(-\Omega_2 - D)}_{E_{\textrm{cc}}^{\textrm{hh-RPA}}},
    \end{equation}
    which is therefore also reproduced by the physical solution of ladder CCD.
    Since \cref{eq:pphhRPA_eqn} is a symplectic eigenvalue problem, we define
    \begin{equation}
        \bar{T}
            = -\begin{bmatrix}
                0 & Y_1 \\
                Y_2^* & 0
            \end{bmatrix}
            \begin{bmatrix}
                X_2^* & 0 \\
                0 & X_1
            \end{bmatrix}^{-1}
            = \begin{bmatrix}
                0 & T_1 \\
                T_1^{\top} & 0
            \end{bmatrix}
    \end{equation}
    which admits an Autonne-Takagi decomposition similar to Eq.~\fakeref{M13}.
    We thus conclude that $\lambda_{\textrm{max}}(\bar{T}^{\dagger} \bar{T}) < 1$ is a sufficient and necessary condition for the physical solution.
    Noting that
    \begin{equation}
        \bar{T}^{\dagger} \bar{T}
            = \begin{bmatrix}
                T_1^* T_1^{\top} & 0 \\
                0 & T_1^{\dagger} T_1
            \end{bmatrix},
    \end{equation}
    the condition for physical pp-RPA solution can be further simplified to be
    \begin{equation}
        \lambda_{\textrm{max}}(T_1^{\dagger} T_1) = \lambda_{\textrm{max}}(T_1^* T_1^{\top}) < 1.
    \end{equation}

    \section{Extension to unstable mean-field references}

    The derivation in the main text for ph-RPA and in \cref{sec:extension_other_flavor} for other flavors of RPA assumes a stable mean-field reference (HF for ph-RPA and pp-RPA and HFB for qp-RPA), which guarantees that all RPA eigenvalues are real.
    However, a stable mean-field reference is not a necessary condition for real RPA frequencies;
    see e.g.,~the Appendix of Ref~\onlinecite{Scuseria13JCP} for a counterexample.
    In this section, we show that all results derived in the main text and in \cref{sec:extension_other_flavor} are valid as long as the RPA frequencies are real, regardless of stability of the underlying mean-field reference.

    For an unstable mean-field reference with real RPA eigenvalues, the RPA eigenvalue equation (Eq.~\fakeref{M2}) still holds.
    However, in this case, $\Omega$ contains both positive and negative RPA frequencies.
    Let
    \begin{equation}
        V_1
            = \begin{bmatrix}
                X \\ Y
            \end{bmatrix},
        \qquad{}
        V_2
            = \begin{bmatrix}
                Y^* \\ X^*
            \end{bmatrix}
            = \begin{bmatrix}
                0 & 1 \\ 1 & 0
            \end{bmatrix} V_1^*
    \end{equation}
    be the eigenvectors corresponding to $\Omega$ and $-\Omega$.
    We follow Ref~\onlinecite{Scuseria13JCP} and define $\Omega$ to be those RPA frequencies whose corresponding eigenvector has a \textit{positive norm}, i.e.,
    \begin{equation}
        V_1^{\dagger} \eta V_1
            = X^{\dagger} X - Y^{\dagger} Y
            = 1,
    \end{equation}
    which automatically implies that $V_2$ has a negative norm,
    \begin{equation}
        V_2^{\dagger} \eta V_2
            = -(V_1^{\dagger} \eta V_1)^*
            = -1,
    \end{equation}
    where the metric matrix is
    \begin{equation}
        \eta
            = \begin{bmatrix}
                1 & 0 \\
                0 & -1
            \end{bmatrix}.
    \end{equation}
    With this definition, the RPA correlation energy is formally the same as Eq.~\fakeref{M1}, i.e.,
    \begin{equation}
        E_{\textrm{c}}^{\textrm{RPA}}
            = \frac{1}{2} \textrm{Tr}\, (\Omega - A).
    \end{equation}
    But again, $\Omega$ now contains some negative frequencies.
    This RPA correlation energy is reproduced by the Riccati equation (Eq.~\fakeref{M7}) with the physical solution $T = Y X^{-1}$, where $X$ and $Y$ are from $V_1$ with positive norm.
    The family of Riccati solutions are then $T_{\bm{\eta}} = Y_{\bm{\eta}} X_{\bm{\eta}}^{-1}$, where $\eta_n = \pm 1$ now corresponds to taking the $n$-th column from $V_1$ (which has a positive norm) and $V_2$ (which has a negative norm), respectively.
    With this modified definition of physical and unphysical solutions, all other conclusions derived in the main text and in \cref{sec:extension_other_flavor} naturally apply to this special case of unstable mean-field references.

    \section{Supplementary data}
    \label{sec:supp_data}

    The following GitHub repository
    \begin{center}
        \url{https://github.com/hongzhouye/supporting_data/tree/main/2025/RPA_Stability}
    \end{center}
    collects (i) optimized molecular structures and (ii) data presented in Fig.~\fakeref{M1} and \fakeref{M2}.

    \section{Supplementary figures}

    \begin{figure}[!h]
        \centering
        \includegraphics[width=6.4in]{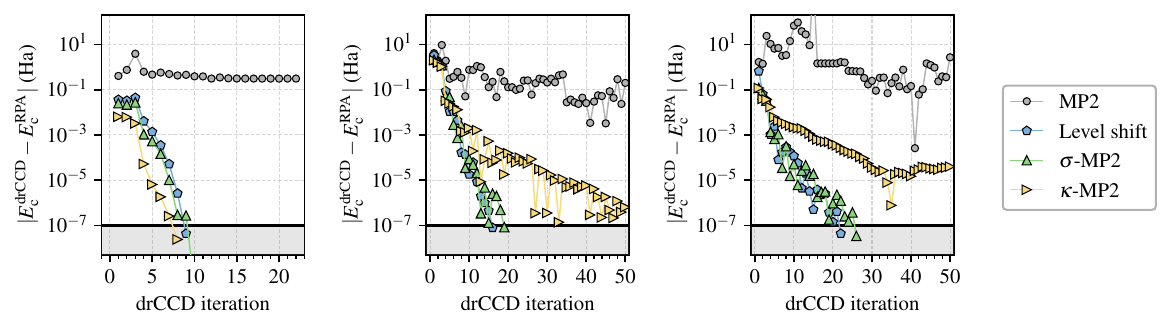}
        \caption{The same figures as Fig.~\fakeref{M1}A, C, and E except that the two-stage algorithm is not used.}
        \label{fig:precond_conv}
    \end{figure}

    \clearpage

    \bibliography{refs}